\documentclass[nobalancelastpage,twocolumn,prl,superscriptaddress]{revtex4-1}

\usepackage{graphicx}

\newcommand{\ket}[1]{\vert#1\rangle}

\usepackage{color}

\begin{document}

\title{Matrix product state approximations for infinite systems}

\author{Norbert Schuch}
\affiliation{Max-Planck-Institute of Quantum Optics, 
Hans-Kopfermann-Str.\ 1, 85748 Garching, Germany}
\author{Frank Verstraete}
\affiliation{Ghent University, Department of Physics and Astronomy,
Krijgslaan 281-S9, 9000 Gent, Belgium}
\affiliation{Vienna Center for Quantum Science, Universität Wien,
Boltzmanngasse 5, 1090 Wien, Austria}

\begin{abstract}
We prove that ground states of gapped local Hamiltonians on an infinite
spin chain can be efficiently approximated by matrix product states
with a bond dimension which scales as $D\sim (\ell-1)/\epsilon$, where
any local quantity on $\ell$ consecutive spins is approximated to accuracy
$\epsilon$.
\end{abstract}

\maketitle

Matrix product states (MPS) provide a systematic way of parameterizing ground
states of local quantum spin chains. In the case of a quantum Hamiltonian
defined on a finite spin chain with $N$ sites and a gap $\Delta$ above the
ground state $\ket{\Phi}$, it has been proven that there exists a matrix
product state $|\psi_D\rangle$ with bond dimension $D$ which approximates
the ground state up to precision $\epsilon=\|\ket{\psi_D}-\ket\Phi\|_2^2$
with
$D=O(\mathrm{poly}(N,1/\epsilon))$~\cite{hastings:arealaw,verstraete:faithfully}.
This fact provides the theoretical justification for the success of the
density matrix renormalization group method.

It has for long been an open problem whether a similar result holds in the
thermodynamic limit (i.e., $N\rightarrow\infty$), or more generally
whether there exist approximations whose cost does not scale with $N$. For the
exisiting results, such a scaling is unavoidable since we are aiming for 
good \emph{global} approximations, yet states necessarily become
orthogonal in the thermodynamic limit.  However, this is clearly asking
too much:  For a faithful approximation of ground states of local
Hamiltonians, a good approximation of \emph{local} quantities is surely 
sufficient.

In this paper, we show how to construct an efficient matrix product
approximation to ground states of local Hamiltonians in the thermodynamic
limit which reproduces all local quantities faithfully.  More precisely,
we will show that in order to approximate the exact ground state of an
infinite spin chain with nearest neighbor
interactions by an MPS with bond dimension
$D$ such that all reduced density matrices $\rho_\ell$ on $\ell$ spins are
$\epsilon$-close in trace norm to the ones of the exact ground state,
$\sigma_\ell$, we only need a bond dimension which scales as 
\[
D(\epsilon)\le
    \frac{6(\ell-1)}{\epsilon}
    \exp\left[\frac{2}{c_2^{3/4}\Delta^{1/4}}
	\left(\log\frac{(\ell-1)\tilde c_1}{\epsilon^3}\right)^{3/4}
    \right]
\]
with $\Delta$ the gap of the infinite spin chain (with the convention that
all local terms have norm $\le1$) and $c_1$ and $c_2$ universal
constants which only depend on the local Hilbert space dimension.

\begin{figure}[t!]
\includegraphics[width=0.9\columnwidth]{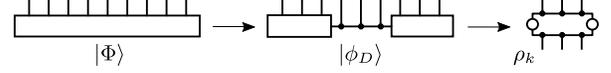}
\caption{\label{fig}
Construction of the MPS. First, $\ket\Phi$ is cut in $k+1$ places to
bond dimension $\chi$, yielding a total error
$\|\ket\Phi-\ket{\phi_D}\|_2^2\le 2\delta (k+1)$~\cite{verstraete:faithfully}.
Then, the remaining spins are traced, yielding an MPO with
$D=\chi^2$ and 
$\|\rho_k-\sigma_k\|_\mathrm{tr}\le2 \sqrt{2\delta (k+1)}$.
}
\end{figure}

This is proven by making use of the results in
Refs.~\cite{arad:rg-algorithms-and-area-laws,huang:area-law}, which show
that the sum $\delta$ of all but the leading $\chi$ Schmidt coefficients
squared for any spin system along any cut is bounded by
\begin{equation}
\label{eq:delta-bound}
\log(\delta)\le\log(c_1)-c_2\Delta^{1/3}\left(\log\chi\right)^{4/3}\ .
\end{equation}
We now construct a matrix product operator (MPO) on $k$ sites by cutting
the exact ground state $k+1$ times and then tracing the outer spins,
Fig.~\ref{fig}. In order to obtain a translational invariant
MPO for the infinite chain, we take an infinite tensor product of this
MPO with itself, and sum over all $k$ translations. The resulting MPO has bond
dimension $D=k\chi^2$, and its reduced state on $\ell$ contiguous sites
approximates the ground state up to error
\[
\epsilon\le 
    \frac{k-(\ell-1)}{k}\|\rho_\ell-\sigma_\ell\|_\mathrm{tr}
    +\frac{2(\ell-1)}{k}
    \le 2\sqrt{2k\delta}+\frac{2(\ell-1)}{k}\ .
\]
Choosing $k=2^{1/3}(\ell-1)^{2/3}/\delta^{1/3}$, we can solve for $\delta$ and
$\chi$, substitute in Eq.~(\ref{eq:delta-bound}), and immediately obtain the
result (with $\tilde c_1=108\,c_1$).

For small $\epsilon$, the exponential term grows slower than
$(\ell/\epsilon)^\alpha$ for any $\alpha>0$ and thus becomes negligible, so
that we effectively arrive at the result that 
\[
D(\epsilon)=O\left(\frac{\ell-1}{\epsilon}\right)\ .
\]
This is an extremely strong result: It shows that all local quantities in
a ground state of any gapped quantum spin chain can be approximated
efficiently using a MPS, at a cost which only scales linearly in the
required precision, and is asymptotically independent of the gap.

\emph{Acknowledgements.---}%
We acknowledge helpful discussions with Y.~Huang and U.~Vazirani.  This
project has been supported by the ERC grants WASCOSYS (No.~636201) and
QUTE (No.~647905).

\end{document}